\documentclass[12pt]{article}
\usepackage{amsfonts,amsmath,amssymb}
\usepackage[dvips]{graphics}
\newcommand{\Rl}{\mathbb{R}}

\newcommand{\Ir}{\mathbb{Z}}
\newcommand{\Cx}{\mathbb{C}}

\newcommand{\B}{\mathcal{B}}

\newcommand{\T}{\mathcal{T}}

\newcommand{\E}{\mathcal{E}}
\newtheorem{theorem}{Theorem}[section]

\newtheorem{lemma}[theorem]{Lemma}
\newtheorem{proposition}[theorem]{Proposition}
\newtheorem{corollary}[theorem]{Corollary}

\newtheorem{conjecture}[theorem]{Conjecture}

\newcommand{\eq}[1]{(\ref{#1})}
\def\QED{{\hspace*{\fill}{$\square$}}\quad
          \vspace{10pt}}
\newenvironment{proof}{\noindent {\bf Proof: }}{\QED\medskip}
\newenvironment{proofof}{\noindent {\bf Proof of }}{\QED\medskip}

\def\idty{{\mathchoice {\rm 1\mskip-4mu l} {\rm 1\mskip-4mu l} %
{\rm 1\mskip-4.5mu l} {\rm 1\mskip-5mu l}}}
\newcommand{\be}{\begin{equation}}
\newcommand{\ee}{\end{equation}}
\newcommand{\bea}{\begin{eqnarray}}
\newcommand{\eea}{\end{eqnarray}}
\newcommand{\beann}{\begin{eqnarray*}}
\newcommand{\eeann}{\end{eqnarray*}}

\newcommand{\ket}[1]{\left\vert #1\right\rangle}

\DeclareMathAlphabet{\mathol}{OT1}{cmr}{l}{ol}

\newcommand{\floor}[1]{\lfloor{#1}\rfloor}
\newcommand{\ip}[2]{\langle{#1|#2}\rangle}

\newcommand{\spec}{\operatorname{spec}}

\newcommand{\Bdd}{\mathcal{B}}
\newcommand{\Hil}{\mathcal{H}}
\newcommand{\Obs}{\mathcal{A}}

\begin{document}
{\baselineskip=10pt \thispagestyle{empty}


\begin{center}
{\Large \bf Ferromagnetic Ordering of Energy Levels\\[27pt]}
{\bf Bruno Nachtergaele, Wolfgang Spitzer\\[0pt]}
{Department of Mathematics\\
University of California, Davis\\
Davis, CA 95616-8633, USA\\[5pt]}
{bxn@math.ucdavis.edu, spitzer@math.ucdavis.edu}\\[15pt]
{\bf Shannon Starr\\[0pt]}
{Department of Physics\\
Princeton University\\
Princeton, NJ 08544, USA\\[5pt]}
{sstarr@math.princeton.edu}\\[15pt]
{\it Dedicated to Elliott Lieb on the occasion of his seventieth birthday}
\end{center}

\noindent
{\bf Abstract}
We study a natural conjecture regarding ferromagnetic ordering of energy levels
in the Heisenberg model which complements the Lieb-Mattis Theorem of 1962 for
antiferromagnets: for ferromagnetic Heisenberg models the lowest energies in
each subspace of fixed total spin are strictly ordered according to the total
spin, with the lowest, i.e., the ground state, belonging to  the maximal total
spin subspace. Our main result is a proof of this conjecture for the spin-1/2
Heisenberg XXX and XXZ ferromagnets in one dimension.  Our proof has two main
ingredients. The first is an extension of a result of Koma and Nachtergaele
which shows that monotonicity as a function of the total spin follows from the
monotonicity of the ground state energy in each total spin subspace as a
function of the length of the chain. For the second part of the proof we use
the Temperley-Lieb algebra to calculate, in a suitable basis, the matrix
elements of the Hamiltonian  restricted to each subspace of the highest weight
vectors with a  given total spin. We then show that the positivity properties
of these matrix elements imply the necessary monotonicity in the volume.
Our method also shows that the first excited state of the XXX
ferromagnet on any finite tree has one less than maximal total spin.
\vspace{5pt}

\newline
{\small \bf MCS numbers:} 82B10, 82B24, 82D40
\vfill
\hrule width2truein \smallskip {\baselineskip=10pt \noindent Copyright
\copyright\ 2003 by the authors. Reproduction of this article in its entirety
is permitted for non-commercial purposes.}}

\newpage

\section{Introduction}

Given any finite set $\Lambda$, and a set of coupling constants
$$
J = \{J_{\{x,y\}}\, :\, \{x,y\} \subset \Lambda\, ,\ x\neq y\}
$$
one defines a Heisenberg model
by specifying a Hamiltonian of the following form:
\begin{equation}
H_{\Lambda,J} = \sum_{\substack{\{x,y\} \subset \Lambda \\ x\neq y}}
      J_{\{x,y\}} \boldsymbol{S}_x \cdot \boldsymbol{S}_y\,.
\end{equation}
Here $\boldsymbol{S}_x = (S_x^1,S_x^2,S_x^3)$ is the defining spin
vector for an irreducible representation of $\textrm{SU}(2)$ at
the site $x \in \Lambda$. In general, the magnitude of the spin at
site $x$ is $s_x\in\frac{1}{2}\mathbb N$.

The Hamiltonian is clearly invariant with respect to the action
of $\textrm{SU}(2)$ on
$\Hil(\Lambda) = \bigotimes_{x \in \Lambda} \Cx^{2s_x+1}$.
Therefore the vectors of a given total
spin (a vector has total spin $S$ if it is an eigenvector
of the Casimir operator
$\sum_{x,y \in \Lambda} \boldsymbol{S}_x \cdot \boldsymbol{S}_y$
with eigenvalue $S(S+1)$)
form an invariant subspace for $H_\Lambda$.
We define $E(\Lambda,J,S)$ to
be the lowest energy among all eigenvectors with total spin $S$.

We say $(\Lambda,J)$ is reducible if there is a proper subset
$\Lambda_1$ such that $J_{\{x,y\}} = 0$ whenever $x\in \Lambda_1$
and $y \in \Lambda \setminus \Lambda_1$, or vice versa.
Otherwise, we call the model \emph{irreducible}.
It is irreducible models which interest us.

If there are two subsets $A,B$ such that $\Lambda = A \sqcup B$
and
$$
\begin{cases}
J_{\{x,y\}} \geq 0 & \textrm{if}\quad x\in A,\, y\in B\quad
     \textrm{or}\quad x\in B,\, y\in A\, ,\\
J_{\{x,y\}} \leq 0 & \textrm{if}\quad x,y \in A\quad
     \textrm{or}\quad x,y\in B,
\end{cases}
$$
we call the model \emph{$(A,B)$-bipartite}.
Such models form a special class.
It makes sense to then define
$S_A = \sum_{x \in A} s_x$ and $S_B = \sum_{x \in B} s_x$.

The Lieb-Mattis Theorem says the following.

\begin{theorem}{(Ordering of Energy Levels \cite{LM})}
\label{thm:LM}\\
Suppose the Heisenberg Hamiltonian
$H_{\Lambda,J}$ is irreducible and $(A,B)$-bipartite.
Define $\mathcal{S} = |S_A - S_B|$.
Then
\begin{align}
&E(\Lambda,S+1) > E(\Lambda,S) \quad \textrm{for all}\quad S\geq \mathcal{S}\, ,\\
&E(\Lambda,S) > E(\Lambda,\mathcal{S}) \quad \textrm{for} \quad S<\mathcal{S}\, .
\end{align}
\end{theorem}

The Lieb-Mattis theorem is simple and elegant, and we repeat the
basic argument, here. The main tool is the Perron-Frobenius
theorem. We quote the Perron-Frobenius theorem from \cite{Simon}
where a proof can be found (pages 130--132).
\begin{theorem}{(Perron-Frobenius)}
If $A = (a_{ij})$ is a square matrix of size $n>1$ with
non-negative entries and such that for some $k\geq 1$, $A^k$ has
strictly positive entries, then
\begin{enumerate}
\item $\rho(A) = \max_{\lambda \in \spec(A)} |\lambda |$ is an
eigenvalue of $A$. \item $\rho(A)$ is simple (in the strong sense
that it is a simple root of $\det(A-\lambda)=0$).
\item For any
other eigenvalue $\lambda$, $|\lambda|<\rho(A)$.
\item The
eigenvector $v$ associated to $\rho(A)$ has only strictly positive
components.
\item No other eigenvector has only non-negative
components.
\end{enumerate}
\end{theorem}

\begin{proofof}{\bf Theorem \ref{thm:LM}:} (This is only a sketch.
See \cite{LM} for details.)
Let $\{\ket{\sigma} : \sigma = (\sigma_x)_{x\in \Lambda}, \sigma_x \in [-s_x,s_x]\}$
be the standard Ising basis of $\Hil(\Lambda)$.
Define $\phi(\sigma) = e^{i(\pi/2)\sum_{x \in A} \sigma_x} \ket{\sigma}$.
In this basis, $H_{\Lambda,J}$ has all real, non-positive off-diagonal entries.
Moreover, since it is assumed to be irreducible, this means that restricted
to each total $S^3$-eigenspace, the matrix representation is irreducible.
Hence, in each $S^3$-eigenspace, the minimum energy vector is unique.
Let $S(\Lambda,J,M)$ be the total spin of the minimum energy vector for
$H_{\Lambda,J}$
in the $S^3$-eigenspace with eigenvalue $M$
(henceforth called the $M$-subspace).

Note that the set of all $J$ such that $H_{\Lambda,J}$ is $(A,B)$-bipartite
forms a convex region of $\Rl^{|\Lambda|(|\Lambda|-1)/2}$.
Hence, it is connected.
Clearly, $S(\Lambda,J,M)$ is a continuous, integer-valued function on this
region for each $M$; therefore, it is constant.
One particular model which is solvable is
$$
J_{\{x,y\}} =
     \begin{cases} -1 & x\in A,\, y\in B\quad \textrm{or}\quad
                        y\in A,\, x\in B\, ;\\
                    0 & x,y \in A\quad \textrm{or}\quad x,y \in B\, .
     \end{cases}
$$
For this model, it is easily seen that
$$
S(J,M) = \begin{cases} |M| & |M|>\mathcal{S}\, ,\\
                       \mathcal{S} & |m|\leq\mathcal{S}\, .
         \end{cases}
$$
This, along with the constancy of $S(\Lambda,J,M)$
for $J$ in the convex set, implies the result.
\end{proofof}

There are three natural categories for $(A,B)$-bipartite
Hamiltonians,
\begin{itemize}
\item antiferromagnetic if $\mathcal{S}=0$;
\item ferrimagnetic if $0<\mathcal{S}<\max(S_A,S_B)$;
\item ferromagnetic if $\mathcal{S} = \max(S_A,S_B)>0$.
\end{itemize}
Note that for antiferromagnets, the Lieb-Mattis theorem
implies
\begin{equation}
E(\Lambda,J,S) < E(\Lambda,J,S')\quad \textrm{whenever}\quad
  S<S'\, .
\end{equation}
The Lieb-Mattis theorem also implies ``ferromagnetic ordering
of the ground state''.
I.e., for ferromagnetic Hamiltonians, the ground state has
maximum possible spin.
A natural guess is that for any irreducible, ferromagnetic model,
\begin{equation}
\label{eq:FerrOrd}
E(\Lambda,J,S) > E(\Lambda,J,S') \quad \textrm{whenever}\quad S<S'\, .
\end{equation}
We call this ``ferromagnetic ordering of energy levels''.

\begin{conjecture}\label{con:main}
For any irreducible, ferromagnetic Heisenberg model there is
ferromagnetic ordering of energy levels.
I.e., (\ref{eq:FerrOrd}) is verified.
\end{conjecture}

In the case of antiferromagnets, the Lieb-Mattis theorem proves
full ordering precisely because the dispersion relation for the
ground state energy in each $M$ subspace, versus $M$, is not flat;
it is increasing in $|M|$. This is crucial because the
Perron-Frobenius theorem only gives direct information about the
ground state in each irreducible sector, and for irreducible
Heisenberg models, the $M$-subspaces are the irreducible sectors.
The fact that, for the ferromagnet, the dispersion relation is
flat proves ferromagnetic ordering of the ground state, but no
more. It is not obvious how to prove Conjecture \ref{con:main}, in
general, though we believe it is true.

We mention a somewhat related difficulty for the ferromagnetic
Heisenberg model: the fact that it is not reflection positive
\cite{Speer}. Reflection positivity is a particular property which
is valid for the Heisenberg antiferromagnet, and in fact the proof of
the Lieb-Mattis theorem for the antiferromagnet can be
considered as an early forerunner of reflection positivity. By
using reflection positivity, Dyson, Lieb, and Simon were able to
prove that the antiferromagnet has a phase transition, at (small)
positive temperatures, in dimensions $d\geq 3$
\cite{DLS}\footnote{The originators of the reflection positivity
approach to proving continuous symmetry breaking  were
\cite{FSS}.}. Jord\~{a}o-Neves and Fernando Perez first used
reflection positivity to prove a phase transition for the
Heisenberg antiferromagnet in two dimensions for $s_x\geq 3/2$
\cite{NP}. The analogous result for $d=2$ and $s_x \geq 1$ as well
as $d\geq 3$ and $s_x \geq 1/2$ was subsequently proved by
Kennedy, Lieb and Shastry \cite{KLS}. Many interesting results on
a variety of topics later followed using reflection positivity
\cite{KLS2,L,LN,MN,LS}. However this technique never succeeded to
prove a phase transition,  at positive temperatures, for the
ferromagnetic Heisenberg model, despite the fact that it is
completely trivial to prove a phase transition for the ground
states. This is simply because the ferromagnetic Heisenberg model
is \emph{not} reflection positive\footnote{For more about this
important problem, see the IAMP  website
\mbox{http://www.math.princeton.edu/\,$\tilde{}$\,aizenman/OpenProblems.iamp/}.
}. Because of this connection, the question of proving
ferromagnetic ordering of  energy levels seems even more
interesting.

We would like to argue that the quantum Heisenberg ferromagnet is
just as interesting as the antiferromagnet. The latter has
received much more attention because its ground state is a highly
non-trivial object, while the same cannot be said of the rather
trivial ground states of the ferromagnet. The situation changes
dramatically, however, when one focuses on the excitation
spectrum, or even just asks for the lowest energy states in
invariant subspaces. E.g., Dhar and Shastry studied the lowest
excited states in the subspaces of fixed momentum \cite{DS}. In
\cite{NSt} two of us determined the ground states in subspaces of
fixed third component of the spin subject to ``droplet'' boundary
conditions. In the present paper we consider ground states in the
subspaces of fixed total spin. In each case, the ferromagnetic
model shows interesting structure.

As a step in the direction of ferromagnetic ordering, Koma and
Nachtergaele~\cite{KN1} proved, for the case of the spin-$1/2$
ferromagnetic Heisenberg spin chain of length $L$, that the lowest
excitation above the ground state is a 1-spin deviate vector,
i.e., with total spin $S=L/2-1$. Thus, for any $S<L/2-1$,
$E([1,L],S)>E([1,L],L/2-1)$. More generally, we will call an
$n$-spin deviate any vector with total spin equal to $L/2-n$.
Their proof involves a very simple argument just using addition of
angular momentum for the Lie group $\textrm{SU}(2)$. Moreover, it
generalizes to the $\textrm{SU}_q(2)$ symmetric XXZ model with
Ising-like anisotropy. Their basic theorem implies that, for any
$L_0$ and $n_0$,  the minimum energy of all $m$-spin-deviates is
less than the minimum energy of all $n$-spin-deviates, for $m\leq
n\leq n_0$ and chains of length $L \leq L_0$, as long as the
minimum energy of any $n$-spin-deviate, with $n\leq n_0$,  is
nonincreasing in $L$ for $L\leq L_0$. Hence, they were able to
calculate the exact spectral gap above the ground states of the
ferromagnetic XXZ model for $s=1/2$ and $d=1$, because they could
completely diagonalize the Hamiltonian restricted to $0$- and
$1$-spin-deviates.

In the present paper, we will reconsider their basic theorem, and show how it
can be generalized to provide information on the ordering of energy levels of
$s=1/2$ ferromagnets.  In particular, we use the theorem to prove complete
ferromagnetic ordering of energy levels for the XXZ and XXX models for which
Koma and Nachtergaele calculated the spectral gap. The Koma-Nachtergaele
theorem is only one piece of the puzzle however. The other piece is an
inequality for the lowest eigenvalues of (not necessarily symmetric) matrices
with non-positive off-diagonal matrix elements. See Lemma \ref{lem:second}.

Loosely stated, the lemma says that if $B$ is an $n\times n$ matrix
with non-positive off-diagonal entries, and $A$ is a $m\times m$
submatrix obtained from $B$ by restricting the range of the indices to
$m$, then the smallest eigenvalue of $B$ is less or equal to the smallest
eigenvalue of $A$.

We apply this lemma to the matrices of the one-dimensional XXX and XXZ models
with respect to the generalized Hulth\'en basis introduced by Temperley and
Lieb \cite{TL}. Indeed, the nearest-neighbor interactions of the XXZ
ferromagnet are generators of the Temperley-Lieb algebra, which is of key
importance.

\begin{theorem}\label{thm:main}
Ferromagnetic ordering of energy levels holds for the spin 1/2
ferromagnetic XXZ chain, of arbitrary length, $L\geq 2$, and anisotropy,
$\Delta \geq 1$.
\end{theorem}

Although the result may seem rather special to the case of the Bethe-Ansatz
solvable XXZ model, it is not really the case. In particular, we also use the
same argument to prove that for the XXX model on any finite tree, the first
excitation is a 1-spin deviate, thus generalizing Koma and Nachtergaele's
original spectral gap result to finite trees. This result shows the
applicability of these arguments to non-integrable spin systems, and may also
be of interest to probabilists since it proves that for any tree, the spectral
gap of the symmetric, simple exclusion process equals the spectral gap of
the random walk.

We believe that our theorems for these particular examples give
credible evidence to Conjecture \ref{con:main}.

\section{Definition of the XXZ model with kink boundary fields}

Our main results regard the spin-$1/2$ XXZ model for anisotropies
$\Delta \in [1,\infty]$. This is a nearest-neighbor Hamiltonian,
\be
H_{[1,L]} = \sum_{x=1}^{L-1} h_{x,x+1}
\label{ham}\ee
with nearest-neighbor interaction
\begin{equation}
h_{x,x+1} = j^2 - S_x^3 S_{x+1}^3 - \Delta^{-1}(S_x^{1} S_{x+1}^1
+ S_x^2 S_{x+1}^2) + j \sqrt{1-\Delta^{-2}} (S_x^{3} - S_{x+1}^{3})\, .
\end{equation}
Here, $j=1/2$. In this definition, $\Delta$ is the anisotropy.
$\Delta=1$ gives the isotropic Heisenberg model.
$\Delta=\infty$ is the Ising model with kink boundary conditions.
There is the usual definition of the spin-$1/2$ matrices
$$
S^{1} = \begin{bmatrix}0 & 1/2\\1/2 & 0\end{bmatrix}\, ,\quad
S^{2} = \begin{bmatrix}0 & -i/2\\i/2 & 0\end{bmatrix}\, ,\quad
S^{3} = \begin{bmatrix}1/2 & 0\\0 & -1/2\end{bmatrix}\, ,
$$
and a subscript refers to the site, or tensor factor, where the
spin matrix acts.

The extra boundary field $\frac{1}{2} \sqrt{1-\Delta^{-2}} (S_x^{3} -
S_{x+1}^{3})$ is chosen to allow a quantum group symmetry, but has
some additional nice features even when $j>1/2$, namely that one can determine
all the finite volume ground states in any dimension \cite{ASW,GW}. In
addition, all the infinite volume ground states for the ferromagnetic XXZ (and
XXX) interaction in  one dimension were determined in \cite{KN3}. This last
result is interesting for, among other things, it gives a strong a posteriori
justification of the chosen boundary fields (or their spin-flipped/reflected
images) on a thermodynamic basis (in addition to its obvious algebraic
attraction), as follows: The infinite volume ground states are defined
independently of the boundary fields in the  Hamiltonian. For this model there
is a special property that, restricting any  pure, infinite volume ground state
to the subalgebra of finite volume observables $\Bdd(\Hil([1,L])) \subset
\Obs_0$, where $\Obs_0$ is the quasilocal algebra \footnote{The algebra of
quasilocal observables is  $\Obs_0 = \overline{\cup_{\Lambda \subset \Ir}
\Bdd(\Hil(\Lambda))}$ in which the union is  restricted to finite subsets
$\Lambda \subset \Ir$, and the closure is in operator norm. A ground state is a
state -- normalized, positive functional -- on this algebra which satisfies
local stability. I.e., $\omega$ is a ground state iff for any local observable
$X$, one has $\omega(X^*[H,X]) \geq 0$, which expresses the fact that the
perturbed state $\omega(X^*\dots X)/\omega(X^*X)$ has higher energy than
$\omega$.} one obtains a density matrix whose range is either in the ground
state space of $H_{[1,L]}$, or else is in the ground state space of the
spin-flipped/reflected image of $H_{[1,L]}$.

The ground state space of $H_{[1,L]}$ is defined as the $E=0$ eigenspace, and
it is easy to see that $H_{[1,L]} \geq 0$.

\section{Quantum Group Symmetry}

As mentioned before this Hamiltonian is quantum group symmetric, where the quantum
group is $\textrm{SU}_q(2) = \mathcal{U}_q(\textrm{sl}(2))$, a deformation of
the (universal enveloping algebra for the Lie algebra of the) Lie group
$\textrm{SU}(2)$. The $q$ refers in this case to a real deformation parameter,
specifically $q \in [0,1]$ is the solution of $\Delta = (q+q^{-1})/2$. Because
$q$ is real, the representation theory of
$\textrm{SU}_q(2)$ for $0<q<1$ is so similar to that of $\textrm{SU}(2)$ that
the reader will hardly notice a difference. The most important difference is
that in place of the usual generators  $S^{3}_{[1,L]}$, $S^+_{[1,L]}$ and
$S^-_{[1,L]}$ of the representation of $\textrm{SU}(2)$ on $\Hil([1,L])$, one
has three matrices $S^+_{q,[1,L]}$, $S^-_{q,[1,L]}$ and $S^3_{q,[1,L]}$
\begin{align}
S^3_{q,[1,L]} &= \sum_{x=1}^L S^{3}_x\, ,\\
S^+_{q,[1,L]} &= \sum_{x=1}^L q^{-2\sum_{y=1}^{x-1} S_y^{3}} S_x^+\, ,\\
S^-_{q,[1,L]} &= \sum_{x=1}^L q^{2\sum_{y=x+1}^L S_y^{3}} S_y^-\, .
\end{align}
Of course, $S^{3}_{q,[1,L]}$ is the same as $S^{3}_{[1,L]}$.
All three of these operators commute with $H_{[1,L]}$.

The Clebsch-Gordon series for $\textrm{SU}_q(2)$ is the same as
that for $\textrm{SU}(2)$. In particular there is a unique (up to
isomorphisms) irreducible representation of dimension $d$ for
$d=1,2,3\dots$. As usual, let $j=\frac{1}{2} (d-1)$ be called the
spin. Then, as for $\textrm{SU}(2)$, the number of irreducible
spin $j=L/2-n$ representation of $\textrm{SU}_q(2)$ in
$\Hil([1,L])$ is the same as the number of noncrossing pairings of
$2n$ of the $L$ linearly ordered vertices $\{1,\dots,L\}$ such
that no pairing spans an unpaired vertex~\cite{TL}. Perhaps more
importantly, if $W^{(j)}$ and $W^{(j')}$ are two irreducible
representation of spin $j$ and $j'$ in $\Hil([1,L])$ and
$\Hil([L+1,L+L'])$, then $W^{(j)} \otimes W^{(j')}$ decomposes
into irreducible representations in $\Hil([1,L+L'])$ according to
$W^{(j+j')} \oplus W^{(j+j'-1)} \oplus \dots  \oplus W^{|j'-j|}$.

\section{Reduction to Monotonicity in the Volume}

For each $L\geq 2, n=0,1,\dots,\floor{L/2}$, let $\Hil([1,L],n)$ be the sum
of all irreducible, spin-$[L/2-n]$ representations of
$\textrm{SU}_q(2)$ in $\Hil([1,L])$. Here $\floor{x}$ is the
greatest integer $n$, such that $n\leq x$. These subspaces are
invariant under the action of the Hamiltonian $H_{[1,L]}$ due to
its quantum group symmetry. For the same set of $n$, define
$$
\E(L,n) = \min\{ \ip{\psi}{H_{[1,L]} \psi} :
\psi \in \Hil([1,L],n)\, ,
\|\psi\|=1\}\, .
$$

One can observe the following simple fact, which applies to Hamiltonians
more general than Heisenberg or XXZ models.
\begin{lemma}
\label{lem:IndIneq}
Let $H_{[1,L]}$ and $H_{[1,L+1]}$ be self-adjoint operators
on $\Hil([1,L]) = (\Cx^2)^{\otimes [1,L]}$ and
$\Hil([1,L+1]) = (\Cx^2)^{\otimes [1,L+1]}$, respectively.
Suppose both commute with the action of $SU_q(2)$.
Also,
suppose $H_{[1,L+1]} \geq H_{[1,L]}$,
identifying $H_{[1,L]}$ with the operator on $\Hil([1,L+1])$
obtained by tensoring with the identity on the last factor.
Then for any $n<(L+1)/2$,
\begin{equation}
\label{eq:IndIneq}
     \mathcal{E}(L+1,n) \geq
     \min\{\mathcal{E}(L,n),\mathcal{E}(L,n-1)\}\, ,
\end{equation}
while $E(L+1,(L+1)/2) \geq E(L,(L-1)/2)$.
\end{lemma}

\begin{proof}
By the standard rules of addition of angular momentum,
for any $\psi \in \Hil([1,L+1],n)$, there are four vectors
$\psi_1,\psi_2 \in \Hil([1,L],n)$ and $\psi_3,\psi_4 \in \Hil([1,L],n-1)$,
such that
$$
\psi = \psi_1\otimes\ket{+1/2} + \psi_2\otimes\ket{-1/2}
+ \psi_3\otimes\ket{+1/2} + \psi_4\otimes\ket{-1/2}\, .
$$
Note that $\psi_1\otimes\ket{+1/2},\dots,\psi_4\otimes\ket{-1/2}$
are orthogonal because they are all eigenvectors for the commuting
operators $S^3$ and $\boldsymbol{S}\cdot\boldsymbol{S}$, and for any two vectors, either
the eigenvalues of $S^3$ are different or the eigenvalues of
$\boldsymbol{S}\cdot\boldsymbol{S}$ are different (or both). Moreover, for that same reason
they are also orthogonal with respect to $H_{[1,L]}$. Thus
\begin{align*}
     \frac{\ip{\psi}{H_{[1,L+1]} \psi}}{\|\psi\|^2}
     &\geq \frac{\ip{\psi}{H_{[1,L]} \psi}}{\|\psi\|^2}\\
     &= \sum_{k=1}^4 \frac{\|\psi_k\|^2}
      {\sum_{l=1}^4 \|\psi_l\|^2}\,
      \cdot\, \frac{\ip{\psi_k}{H_{[1,L]} \psi_k}}{\|\psi_k\|^2} \\
     &\geq \min_{k=1,2,3,4} \frac{\ip{\psi_k}{H_{[1,L]} \psi_k}}{\|\psi_k\|^2} \\
     &\geq \min\{\mathcal{E}(L,n), \mathcal{E}(L,n-1)\}\, .
\end{align*}
\end{proof}

The natural generalization to higher spins is immediately obvious.
Suppose each factor in $\Hil([1,L]) = (\Cx^{2j+1})^{\otimes
[1,L]}$ and $\Hil([1,L+1]) = (\Cx^{2j+1})^{\otimes [1,L+1]}$ is
canonically equipped with a spin-$j$ representation of
$\textrm{SU}_q(2)$, and that $H_{[1,L]}$ and $H_{[1,L+1]}$ commute
with the actions of $\textrm{SU}_q(2)$ on the products. Defining
$\Hil([1,L],n)$ to be the sum of the spin-$[jL-n]$
representations, we would determine that if $H_{[1,L+1]} \geq
H_{[1,L]}$, then
\begin{equation}
     \mathcal{E}(L+1,n) \geq
     \min_{k=0,\dots,2j+1} \mathcal{E}(L,n-k)\, .
\end{equation}
However, the lemma is most useful as it is stated, for spins-$1/2$,
because of the immediate corollary:

\begin{corollary}
Suppose $H_{[1,L]}$ and $H_{[1,L+1]}$ satisfy the hypotheses of Lemma
\ref{lem:IndIneq}, and suppose further that there is a value $n \in \{0,\dots,\floor{L/2}\}$
such that
\begin{equation}
     \mathcal{E}(L,n)\, <\,
     \mathcal{E}(L,r) \quad \textrm{for all}\quad r>n.
\end{equation}
Then, if $\mathcal{E}(L+1,n) < \mathcal{E}(L,n)$, then also
\begin{equation}
     \mathcal{E}(L+1,n)\, <\,
     \mathcal{E}(L+1,r) \quad \textrm{for all}\quad r>n.
\end{equation}
\end{corollary}

Applying the corollary inductively leads to the following important
result.

\begin{proposition}\label{Prop1}
Suppose that for all $L\in \{2,\dots,L_0\}$, there is defined a
self-adjoint operator $H_{[1,L]}$ on $\Hil([1,L])=(\Cx^2)^{\otimes
[1,L]}$,  commuting with the action of $\textrm{SU}_q(2)$, and
such that $H_{[1,L+1]} - H_{[1,L]}\geq 0$ for $L \in
\{2,\dots,L_0-1\}$. Suppose that, for some $n \in
\{0,\dots,\floor{L_0/2}\}$, $\E(L,n)$ is strictly decreasing as a
function of $L$ for $L \in \{2n,\dots,L_0\}$. Then
\begin{equation}
\mathcal{E}(L_0,n) < \mathcal{E}(L_0,r) \quad \textrm{for all}
  \quad r>n\, .
\end{equation}
\end{proposition}

Here is a more explicit statement of Theorem \ref{thm:main},
expressing ferromagnetic ordering for the spin-$1/2$ XXZ (and XXX)
chains:

\begin{proposition} \label{prop:main}
For the ferromagnetic, spin-$1/2$ XXZ chain, with
$1\leq \Delta < \infty$, the sequence $\E(L,n)$
is strictly increasing in $n$ for $n \in \{0,1,\dots,\floor{L/2}\}$.
\end{proposition}

The proof of this proposition, and thus of Theorem \ref{thm:main},
is obtained by combining Proposition \ref{Prop1} in this section and
Proposition \ref{prop:monotonicity} in Section \ref{sec:monotonicity}.

Although the Bethe Ansatz, in principle, should allow one to diagonalize
$H_{[1,L]}$ in the sectors $\Hil([1,L],n)$, it seems nearly impossible
to extract the required information on the eigenvalues from such an
exact solution even for relatively small $n$. It turns out that it is useful
to reformulate the problem in terms of the following quantities:
$$
\tilde{\E}(L,n) = \min_{r\in\{n,\dots,\floor{L/2}\}}
\E(L,r)\, ,
$$
The sequence $(\tilde{\E}(L,n))_{n\geq 0}$ is the lower, nondecreasing  hull
of the sequence $(\E(L,n))_{n\geq 0}$.

The conclusion
of Proposition \ref{prop:main} is
that  $\tilde{\E}(L_0,n) < \tilde{\E}(L_0,n+1)$.
If one relaxes the hypotheses of Proposition \ref{prop:main}
by allowing non-strict inequalities in place of the strict inequalities,
it is clear what the conclusion will be, and this is equivalent to the
statement that $\tilde{\E}(L_0,n) = \E(L_0,n)$.
We will say that we have proved
\emph{ferromagnetic ordering to level $n$} if we can show that
$\tilde{\E}(L,m) = \E(L,m)$ for $m=1,\dots,n$, and \emph{strict ferromagnetic
ordering to level $n$} if
$$
\tilde{\E}(L,1) < \tilde{\E}(L,2) < \dots < \tilde{\E}(L,n)
< \tilde{\E}(L,n+1)\, .
$$

The property that the ground state subspace has maximal spin is equivalent to
``ferromagnetic ordering to level $0$''; the existence of a non-vanishing
spectral gap above the ground state is equivalent to ``strict ferromagnetic
ordering to level $0$''; the proof that the first excitation lives in the
sector $\Hil([1,L],1)$ implies ``ferromagnetic ordering to level 1''; and the
subsequent proof that the first excitation is minimally degenerate, i.e.,  that
the entire eigenspace is a spin $L/2-1$ irreducible representation, is
proof or ``strict ferromagnetic ordering to level 1''.
Those four results are contained in \cite{LM} and \cite{KN1}.

\section{The Spectral Gap for the XXX and XXZ Spin Chain}

In this section we will show how Proposition \ref{Prop1} can be used to
obtain the spectral gap of the XXZ model on a chain.

\begin{theorem}(Koma and Nachtergaele 1997)
For the spin-$1/2$ XXZ spin chain with $\textrm{SU}_q(2)$ symmetry,
the spectral gap equals
\begin{equation}
\gamma_L = 1 - \Delta^{-1} \cos(\pi/L)\, .
\end{equation}
\end{theorem}

\begin{proof}
By Proposition \ref{Prop1}, it suffices to prove that
$$
\widetilde{\mathcal{E}}(L,1) = 1-\Delta^{-1} \cos(\pi/L)\, ,
$$
because this sequence is decreasing in $L$. We observe that the
quantity $1-\Delta^{-1} \cos(\pi/L)$ is actually the minimum
eigenvalue for the matrix $1-\Delta^{-1} A$ acting on
$\ell^2(\{1,\dots,L-1\})$, where $A$ is the adjacency matrix of
$\{1,\dots,L-1\}$. Namely,
\begin{equation}
A(x,y) = \frac{1}{2} (\delta_{y,x+1} + \delta_{y,x-1})\, .
\end{equation}
This is a clue to the calculation.

Note that by the quantum group symmetry one can calculate
$\E_{L,1}$ by calculating the spectral gap in the
$M=L/2-1$ subspace. In this subspace, we can write
\begin{equation}
\ket{x} = S_x^- \ket{\uparrow}\, ,
\end{equation}
where $\ket{\uparrow}$ is the all-upspin state.
Then we observe
\begin{equation}
     H_{[1,L]} \ket{x}
    = \frac{1}{1+q^2} \sum_{y=1}^L \Big[
    \delta_{y,x-1} (\ket{x} - q \ket{y})
    + \delta_{x,y-1} (q^2 \ket{x} - q \ket{y}) \Big]\, .
\end{equation}
The ground state is proportional to
\begin{equation}
\Psi_0 = \sum_{x=1}^L q^x \ket{x}\, .
\end{equation}

Let us define the \emph{Hulth\'en bracket basis} for the orthogonal complement
of the ground state:
\begin{equation}
\ket{\phi_x} = \ket{x} - q \ket{x-1}
\end{equation}
for $x=2,\dots,L$.
Then observe that
\begin{equation}
     H_{[1,L]} \ket{x}
    = \frac{1}{1+q^2} \sum_{y=1}^L \Big[
    \delta_{y,x-1} \ket{\phi_x}
    - q \delta_{x,y-1} \ket{\phi_y} \Big]\, .
\end{equation}
Hence, if $x\in \{2,\dots,L\}$,
\bea
     &&H_{[1,L]} \ket{\phi_x} \nonumber\\
    &=& \frac{1}{1+q^2} \sum_{y=1}^L \Big[
    \delta_{y,x-1} \ket{\phi_x}
    - q \delta_{x,y-1} \ket{\phi_y}
    - q \delta_{y,x-2} \ket{\phi_{x-1}}
    + q^2 \delta_{x-1,y-1} \ket{\phi_y}\Big] \nonumber\\
    &=& \ket{\phi_x}
    - \frac{q}{1+q^2} \sum_{y=1}^L \Big[
    \delta_{x,y-1} \ket{\phi_y}
    + \delta_{y,x-2} \ket{\phi_{x-1}} \Big]\, .
\eea
Hence in this basis, the representation is precisely
$1-\Delta^{-1} A$, as defined above. So we are done.
\end{proof}

We conclude this section with a few remarks. First of all, the Hulth\'en  basis
has been discovered and rediscovered many times. Although here we have used
just the simplest version, with just one Hulth\'en bracket, one can also obtain
a \emph{non-orthogonal} basis for the highest weight vectors of total spin $j$
subspace using these brackets, which we will use in Section
\ref{sec:monotonicity}. To the best of our knowledge, the first to prove that
the Hulth\'en basis is actually linearly independent were Temperley and Lieb
\cite{TL}. We also refer the reader to their paper for more details about the
basis. In more recent literature, one often finds the term ``Hulth\'en
bracket'' replaced by ``valence bond'', in analogy with chemistry.

A second remark is in order. The calculation of the spectral gap worked so
simply because the representation of the Hamiltonian in the  one-bracket
Hulth\'en basis is actually symmetric. Since the basis is not orthogonal,
there is no reason to expect that to be the case in general. Indeed, if one
considers the two-bracket Hulth\'en basis then the matrix representation is
not orthogonal. As we will show, this is not a serious obstacle as long as
the off-diagonal matrix elements are non-negative.

\section{The Spectral Gap of the XXX Model on a Tree}

Let us consider a sequence of trees $\{T_L\}_{L=2}^{\infty}$
such that $|T_L| = L$ and $T_L$ is the induced subgraph on some $L$
vertices of $T_{L+1}$. We consider the usual XXX ferromagnet
\begin{equation}
     H_{T_L} = \sum_{\{x,y\} \subset T_L\, ,\
    x\sim y} \Big[
    \frac{1}{4} - \boldsymbol{S}_x \cdot \boldsymbol{S}_y \Big]\, .
\end{equation}
We can then prove the following theorem.

\begin{theorem}\label{thm:trees}
One has $\E(L,1) < \E(L,n)$
for any $n>1$.
\end{theorem}

\begin{proof}
To begin the proof, note that Proposition \ref{Prop1} applies
to the set of graphs $\{T_L\}_{L=2}^\infty$ with no changes,
because the Hamiltonian $H_{T_L} \leq H_{T_{L+1}}$, and this
is all that is necessary. Hence one may determine that $\E(L,1) < \E(L,n)$
for any $n>1$ if one can prove that $\E(L,1)$ is strictly decreasing in $L$.

Defining $\ket{x} = S_x^- \ket{\uparrow}$, as before, we again
have for any $x \in T_L$
\begin{equation}
\label{treeHam}
     H_{T_L} \ket{x}
    = \frac{1}{2} \sum_{y} \Big[
    \delta_{y,x-1} (\ket{x} - \ket{y})
    + \delta_{x,y-1} (\ket{x} - \ket{y}) \Big]\,
\end{equation}
but with the proper definition of ``$x-1$'' and ``$y-1$''.
Indeed, let us choose a point $O \in T_2$, to call this the
root. Then for any $T_L$, and any $x \in T_L$ there is a unique
non-backtracking path from $O$ to $x$, because $T_L$ is a tree.
The definition of $x-1$ is that $x-1$ is the immediate predecessor
of $x$ on this path.
Note that it is possible that $x-1=y-1$ for some distinct points
$x$ and $y$ in $T_L$, indeed this will be the case unless the
tree is unary (has no splittings, i.e., is a chain).
Also note that $x$ and $y$ are connected by an edge in $T_L$
iff $x=y-1$ or $y=x-1$, and this is the reason that
(\ref{treeHam}) is correct.

We define the obvious analogue of the one-bracket
Hulth\'en states as
\begin{equation}
\ket{\phi_x} = \ket{x} - \ket{x-1}\, ,
\end{equation}
much as before.
Then, again
\begin{equation}
     H_{T_L} \ket{x}
    = \frac{1}{2} \sum_{y \in T_L} \Big[
    \delta_{y,x-1} \ket{\phi_x}
    - \delta_{x,y-1} \ket{\phi_y} \Big]\, ,
\end{equation}
and
\begin{align}
     H_{T_L} \ket{\phi_x}
    &= \frac{1}{2} \sum_{y\in T_L} \Big[
    \delta_{y,x-1} \ket{\phi_x}
    -  \delta_{x,y-1} \ket{\phi_y}
    -  \delta_{y,x-2} \ket{\phi_{x-1}}
    +  \delta_{x-1,y-1} \ket{\phi_y}\Big] \nonumber\\
    &= \ket{\phi_x}
    - \frac{1}{2} \sum_{y\in T_L} \Big[
    \delta_{x,y-1} \ket{\phi_y}
    + \delta_{y,x-2} \ket{\phi_{x-1}}
    + \delta_{x-1,y-1}(1-\delta_{x,y}) \ket{\phi_y} \Big].
\end{align}
We claim that the matrix $A_L$ defined such that
\begin{equation}
     A_L \ket{\phi_x}
    = \frac{1}{2} \sum_{y \in T_L} \Big[
    \delta_{x,y-1} \ket{\phi_y}
    + \delta_{y,x-2} \ket{\phi_{x-1}}
    + \delta_{x-1,y-1}(1-\delta_{x,y}) \ket{\phi_y} \Big]\, ,
\end{equation}
is actually the adjacency matrix for the line graph of $T_L$,
which we denote $\T_L$.

Here $\T_L$ is the graph constructed from $T_L$ by taking
as a vertex set for $\T_L$ the set of all edges $\{x,x-1\}$
in $T_L$.
Then two distinct vertices are connected in $\T_L$ if the edges
are incident to the same vertex for some vertex in $T_L$.
This happens for edges $\{x,x-1\}$, $\{y,y-1\}$ iff
$x=y-1$, $y=x-1$ or $x-1=y-1$.
Of course if $y=x-1$ then $y-1=x-2$ is in $T_L$ and then
$\ket{\phi_{x-1}} = \ket{\phi_y}$.
Then one does indeed see that
\begin{equation}
    A_L \ket{\phi_x} = \frac{1}{2} \sum_{y \in T_L\setminus \{O\}}
    \chi(\{x,x-1\} \sim \{y,y-1\}) \ket{\phi_y}\, ,
\end{equation}
which is the adjacency matrix (with our $1/2$ normalization). One
particular implication is that the matrix representation for
$H_{T_L}$ in the one-bracket Hulth\'en basis is symmetric.

An important point is that $\T_L$ is the induced subgraph of $\T_{L+1}$,
induced by the edges which lie in $T_L$. Since the matrix $1-A_L$ has
non-positive off-diagonal matrix elements, and since $A_L$ is a  submatrix of
$A_{L+1}$, we can apply Lemma \ref{lem:second}, which  is proved in Section
\ref{sec:monotonicity}. Therefore, this proves the ground state energy (and in
this case actually also the sum of any first $k$ eigenvalues) is nonincreasing
with $L$. Since $T_L$ is connected, it is easy to see, by the Perron-Frobenius
Theorem, that actually the ground state energy is strictly decreasing.
\end{proof}

There is a fruitful connection between Markov processes and
quantum spin systems. The study of the low-lying spectrum for
quantum spin systems relates directly to estimating mixing times
for simple exclusion processes, a subject of continued interest
(see e.g., \cite{Mor} and the references in that paper). In the
Markov process language, this theorem implies that the spectral
gap of the symmetric simple exclusion process equals the spectral
gap of the random walk on any finite tree. This connections
between Markov processes and certain quantum spin models such as
the XXX and XXZ, as well as  closely related models, such as
simple exclusion processes, has been fruitfully exploited in the
past. E.g., in \cite{CM}, Caputo and Martinelli proved good lower
bounds for the gap of the spin-$S$ XXZ chain by mapping the
problem to an asymmetric simple exclusion process and applying
probabilistic techniques to the latter. An example where a similar
relation was used in the other direction is the work of Gwa and
Spohn \cite{GS}. They determine the scaling exponent for the
stationary correlation function of the noisy Burgers equation in
terms of spectral information of an XXX chain obtained through the
Bethe Ansatz.

\section{Monotonicity of the energy}{\label{sec:monotonicity}}

The goal of this section is to prove the following proposition
for the XXZ spin chain.

\begin{proposition}\label{prop:monotonicity}
\be
\E(L+1,n) < \E(L,n),\quad \textrm{for all}\quad L\geq 2\, ,\ n\geq 0\, .
\ee
\end{proposition}

In combination with Proposition \ref{Prop1}, this result provides the
proof for Theorem \ref{thm:main}.

The proof of this proposition relies on two lemmas, Lemma \ref{lem:first} and
Lemma \ref{lem:second}, which we state and prove at the end of this section.
These lemmas are applied to the matrices of the spin-1/2 ferromagnetic
Heisenberg Hamiltonian restricted to the invariant subspaces of all highest
weight vectors of a given total spin.
\vspace{10pt}

\begin{proofof}{\bf Proposition \ref{prop:monotonicity}:}
$\E_{L+1,n}$ is the minimal energy in the subspace with total spin  $L/2-n$.
Clearly, we can restrict the minimization of the energy further to highest
weight vectors in this subspace of fixed total spin, i.e., the eigenvectors of
$S^3$ in $\Hil([1,L],n)$ with eigenvalue $L/2-n$. A convenient basis for this
intersection was introduced by Temperley and Lieb \cite{TL}. They called the
element of this basis {\em generalized Hulth\'en brackets} and proved that
they are linearly independent.

We now apply Lemma \ref{lem:second} with $A=A_{L,n}$ and $B=A_{L+1,n}$,
which satisfy the conditions due to Lemma \ref{lem:first}. The strict
inequality is obtained by the comments following Lemma \ref{lem:second}.
\end{proofof}

\begin{lemma}\label{lem:first}
The square matrices $A_{L,n}$ have the following properties (i) all of their
off-diagonal matrix elements are non-positive, (ii) for all $n$, and $L$,
$A_{L,n}$ is ``embedded'' in $A_{L+1,n}$, in the sense that there is a
subset of the index set of $A_{L+1,n}$, such that $A_{L,n}$ is
the restriction of $A_{L+1,n}$ to that subset.
\end{lemma}

\begin{proof}
Each basis elements is labeled by a configuration of $n$ arcs, each of
which pairs two sites, say $i,j\in\{1,2,\ldots,L\}, i<j$ together, and the
configuration has the properties that arcs are non-crossing and do not span
unpaired sites.
See Figures \ref{fig:hulthen_brackets1}-\ref{fig:hulthen_brackets3},
for a few examples.
We will denote such an arc by $(ij)$, and denote configurations of arcs by
$\alpha, \beta, ...$, and the set of all such configurations for given
$L$ and $n$ by $\B_{L,n}$.

\begin{figure}
\begin{center}
\resizebox{7truecm}{4truecm}{\includegraphics{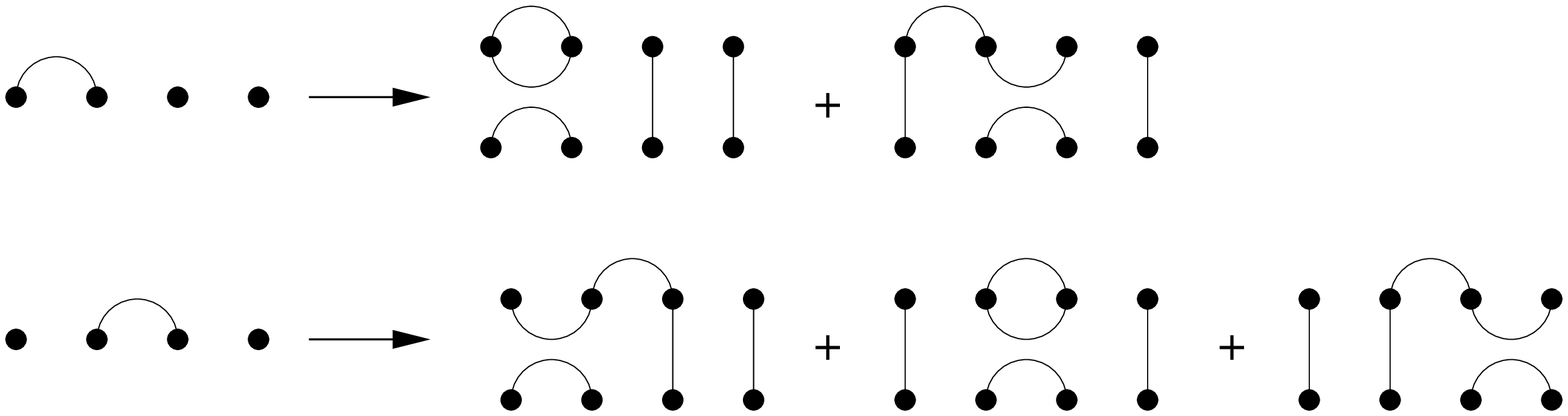}}\qquad
\end{center}
\caption{\label{fig:hulthen_brackets1}
Example of the action of the Hamiltonian of the spin-1/2 XXX or XXZ
chain on a generalized Hulth\'en bracket, for $L=4$, $n=1$.}
\end{figure}

\begin{figure}
\begin{center}
\resizebox{7truecm}{4truecm}{\includegraphics{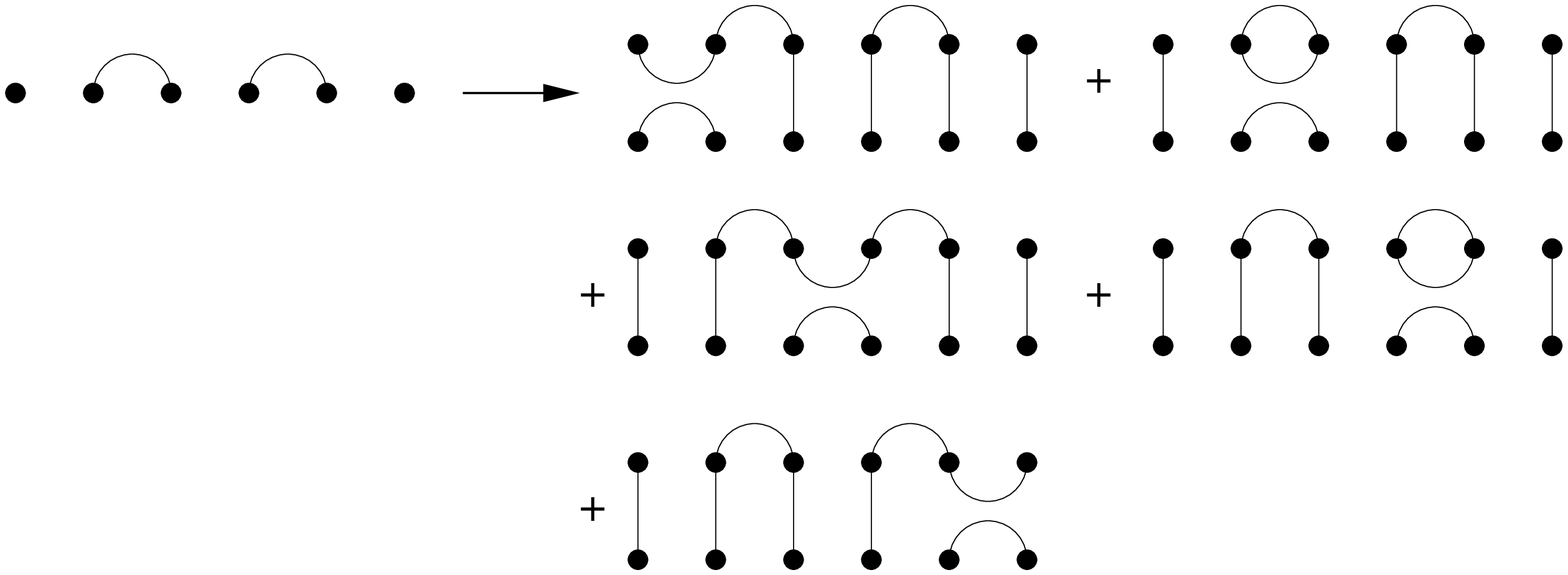}}\qquad
\end{center}
\caption{\label{fig:hulthen_brackets2}
Example of the action of the Hamiltonian of the spin-1/2 XXX or XXZ
chain on a generalized Hulth\'en bracket, for $L=6$, $n=2$.}
\end{figure}

\begin{figure}
\begin{center}
\resizebox{7truecm}{4truecm}{\includegraphics{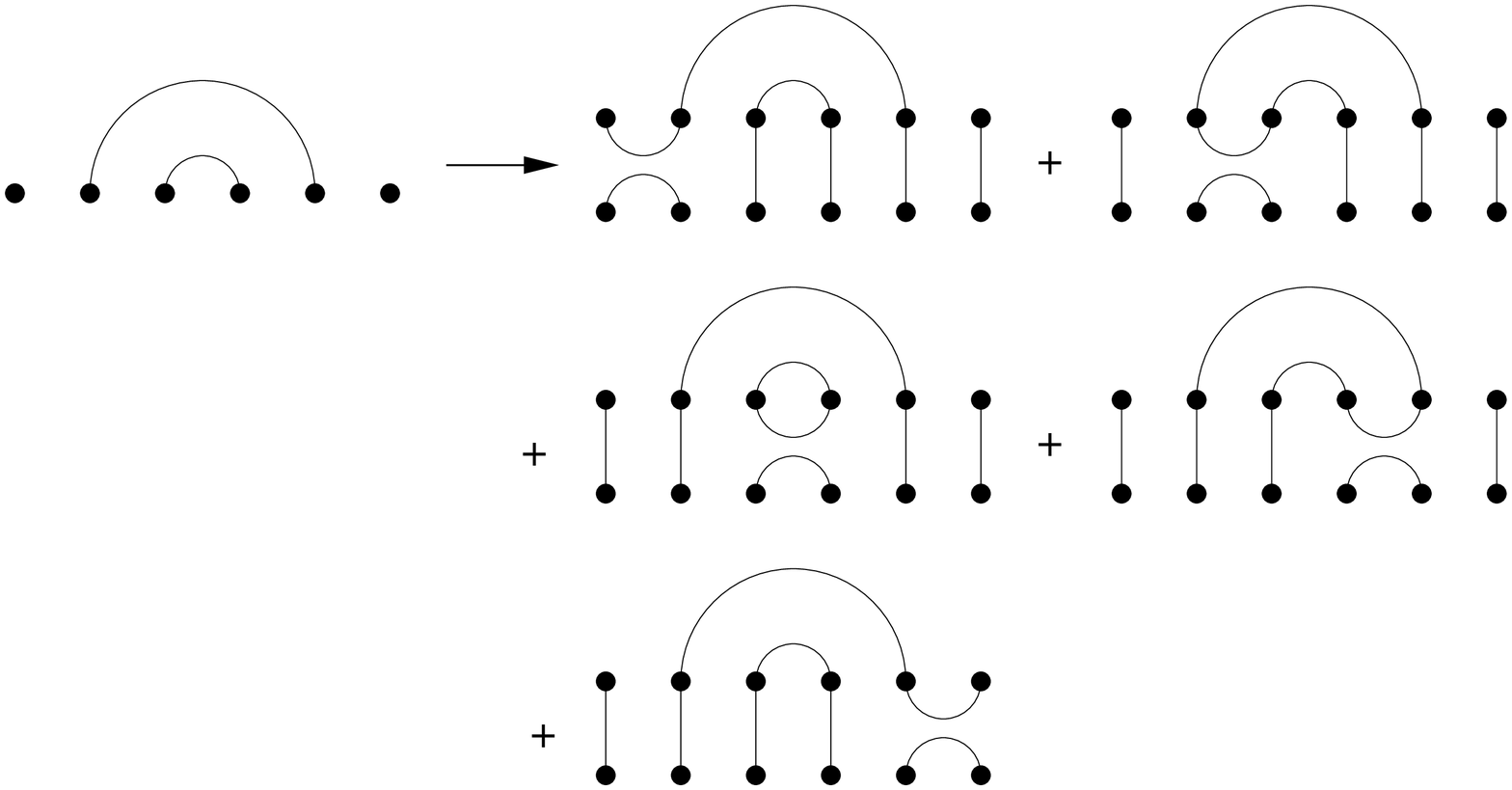}}\qquad
\end{center}
\caption{\label{fig:hulthen_brackets3}
Example of the action of the Hamiltonian of the spin-1/2 XXX or XXZ
chain on a generalized Hulth\'en bracket, for $L=6$, $n=2$.}
\end{figure}

The highest weight vector $\phi_\alpha \in \Hil_L$, corresponding
to the configuration of arcs $\alpha$, are obtained as  tensor
product of the following factors: a factor $\ket{+}$ for each unpaired site,
and a factor $q^{-1/2}\ket{+}_i\ket{-}_j-q^{1/2}\ket{-}_i\ket{+}_j$ for each
arc $(ij)$.

Let $A_{L,n}$ denote the matrix of the Hamiltonian \eq{ham}
with respect to this basis. As the basis is
not orthogonal we should, in general, not expect $A_{L,n}$ to be symmetric.

The matrix elements of $A_{L,n}$ can be most easily computed by also using
a graphical representation of the Hamiltonian, i.e., by writing it in terms
of the generators of the Temperley-Lieb algebra $U_{i,i+1}$, $1\leq i\leq L-1$,
namely
$$
U_{i,i+1}= - (q+q^{-1})h_{i,i+1}\, .
$$

Let $\phi_\alpha$, $\alpha\in\B_{L,n}$, be a basis vector. As
$H_{[1,L]}=-(q+q^{-1}) \sum_{i=1}^{L-1}U_{i,i+1}$, we just have to calculate
$U_{i,i+1}\phi_\alpha$. It turns out that for all $i$ and $\alpha$ there exist
$\beta$ and a real constant $c$ such that $U_{i,i+1}\phi_\alpha=c\phi_\beta$.
The configuration $\beta$ and the constant $c$ are determined by a simple
graphical procedure illustrated in Figures
\ref{fig:hulthen_brackets1}--\ref{fig:hulthen_brackets3}.

We observe the following general rules: (i) if $i$ and $i+1$ are both unpaired
arcs in $\alpha$, we have $U_{i,i+1}\phi_\alpha=0$, (ii) if the composition of
$\alpha$ and $U_{i,i+1}$ is isotopic to $\beta$, with $\beta\neq\alpha$, then
the $U_{i,i+1}\phi_\alpha=\phi_\beta$, i.e., $c=1$, (iii) if $\alpha=\beta$,
$c=- (q+q^{-1})$. This only happens when the ``cup'' of $U_{i,i+1}$ is
paired with an arc in $\alpha$, i.e., $\alpha$ must contain the arc $(i,i+i)$.

With these observations the proof of the lemma is easily completed.
\end{proof}

One can easily use the same observations to explicitly calculate any desired
matrix element, but the properties given in the above lemma are sufficient
for our purposes here.

The next lemma will allow us to compare the smallest eigenvalues of  $A_{L,n}$
and $A_{L+1,n}$. For this it is important that the larger matrix, i.e.,
$B=A_{L+1,n}$, may have positive matrix elements on the diagonal and that there
is only the condition that the first $k$ of those are bounded by the
corresponding diagonal elements of $A$. No assumption about the remaining $l-k$
diagonal elements is made.

\begin{lemma}\label{lem:second}
Let $A=(a_{ij})$ and $B=(b_{ij})$ be two square matrices with real entries
of size $k$ and $l$, respectively, with $l\geq k$, and such that
\beann
&&a_{ij}\leq 0, b_{ij}\leq 0, \mbox{ for all } i\neq j,\\
&&b_{ij}\leq a_{ij}, \mbox{ for } 1\leq i,j\leq k \, .
\eeann
Then
\be
\inf \spec B \leq \inf \spec A\, .
\label{spec-ineq}\ee
\end{lemma}

\begin{proof}
The main idea for the proof of this lemma is taken from Lemma 3.6 in \cite{NS}.
Let $C=\max\{a_{ii}, b_{jj}\mid 1\leq i\leq k, 1\leq j\leq l\}$.
Then, the matrices $\tilde A=C\idty-A$, and $\tilde B=C \idty- B$, have all
non-negative entries, denoted by $\tilde{a}_{ij}$, and $\tilde{b}_{ij}$,
respectively, and $\tilde{b}_{ij}\geq \tilde{a}_{ij}$, for $1\leq i,j \leq k$.

By the Perron-Frobenius Theorem, for any square matrix
$D$ with non-negative entries, and with spectral radius $\rho(D)$, we have that
there is only one eigenvalue with absolute value equal to $\rho(D)$, namely
$\rho(D)$ itself. Let $\lambda_0(M)$ denote the smallest eigenvalue of any
square matrix $M$. Then, by the previous consideration,
$\lambda_0(A)=C-\rho(\tilde A)$ and $\lambda_0(B)=C-\rho(\tilde B)$. Hence, to
prove the lemma, we need to show $\rho(\tilde A)\leq \rho(\tilde B)$.

For any $m\times m$ matrix $D$ with non-negative entries, we also have that
for any $v\in \Cx^m$, $\Vert D v\Vert \leq \Vert D \vert v\vert\Vert$, where
$\vert v\vert$ is the vector with components given by the absolute values of
the components of $v$. Hence,
$$
\Vert D\Vert= \sup_{0\neq v\in \Cx^m}\frac{\Vert D v\Vert}{\Vert v \Vert}
= \sup_{0\neq v\in (\Rl^+)^m}\frac{\Vert D v\Vert}{\Vert v \Vert}\, .
$$

For any $r\geq 1$,
$$
\Vert \tilde{B}^r\Vert
=\sup_{0\neq v\in (\Rl^+)^l}\frac{\Vert \tilde{B}^r v\Vert}{\Vert v \Vert}
\geq \sup_{0\neq w\in (\Rl^+)^k}\frac{\Vert \tilde{B}^r \tilde w\Vert}{\Vert
\tilde w \Vert}\, ,
$$
where, for any $w\in \Cx^k$, we let $\tilde w \in\Cx^l$ denote the vector with
the first $k$ components given by those of $w$,  and the remaining $l-k$
components equal to zero. Clearly, $\Vert \tilde w\Vert =\Vert w\Vert$.

Now, consider $\hat{B}=\tilde{B}_1\oplus\tilde{B}_2$, where $\tilde{B}_1$
is the $k\times k$ matrix with entries $\tilde{b}_{ij}$, $1\leq i,j\leq k$,
and $\tilde{B}_2$ is the diagonal $(l-k)\times(l-k)$ matrix with diagonal
entries $\tilde{b}_{ii}$, $k+1\leq i\leq l$. Then $\tilde{B}\geq \hat{B}$,
elementwise, and hence $\tilde{B}^r\geq \hat{B}^r=
\tilde{B}_1^r\oplus\tilde{B}_2^r$. Moreover, $\tilde{B}_1\geq \tilde{A}$
by the assumptions of the lemma. Therefore, we have
\be
\Vert \tilde{B}^r\Vert\geq
\sup_{0\neq w\in (\Rl^+)^k}\frac{\Vert \tilde{A}^r w\Vert}{\Vert w \Vert}
= \Vert \tilde{A}^r\Vert\, .
\label{power_r}\ee
By taking $r$-th roots and $\limsup$'s, from the
inequality \eq{power_r} we obtain $\rho(\tilde A)\leq \rho(\tilde B)$.
\end{proof}

Sufficient conditions under which the inequality in \eq{spec-ineq}
is strict are easy to find. E.g., when the matrices are
irreducible (in the Perron-Frobenius sense), it is sufficient that
one of the off-diagonal matrix elements $b_{ij}$ of $B$, with at
least one of the indices $i$ or $j>k$. This is the situation in
our application with $A=A_{L,n}$ and $B=A_{L+1,n}$. Another
sufficient condition that guarantees strict inequality in the
irreducible case, is that $b_{ij}<a_{ij}$, for at least one pair
of $i\neq j, 1\leq i,j \leq k$.

\section{Conclusion}

In this paper we have formulated a natural conjecture for ferromagnetic
Heisenberg models, Conjecture \ref{con:main}.
We proved this conjecture for the spin-$1/2$ XXX chain with open
boundary conditions, as well as the analogous results for the
$\textrm{SU}_q(2)$-symmetric spin-$1/2$ XXZ chain.
The techniques developed allow trivial extension to nearest-neighbor
spin chains whose coupling constants $J_{\{x,x+1\}}$ are not all constant,
but are all negative.

To demonstrate the generality of the underlying techniques, we have
also proved that the first excited eigenvector for the XXX model
on a tree is always a 1-spin-deviate.

\section*{Acknowledgements}

Shannon Starr thanks Elliott Lieb for helpful advice on the literature related
to this research. He also gratefully acknowledges an NSF Postdoctoral
Research Fellowship (NSF-MSPRF). This work was supported in part by the
National Science Foundation under Grant \# DMS-0303316. We thank the referees
for pointing out some of the references to us.

\providecommand{\bysame}{\leavevmode\hbox to3em{\hrulefill}\thinspace}


\begin{thebibliography}{10}
\bibitem{ASW} F.C. Alcaraz, S.R. Salinas, and W.F. Wreszinski,
\emph{Anisotropic ferromagnetic quantum domains},
Phys. Rev. Lett \textbf{75} (1995), 930--933.

\bibitem{CM}
P. Caputo and F .Martinelli,
\emph{Relaxation time of anisotropic
simple exclusion processes and quantum Heisenberg models},
Ann. Appl. Probab., {\bf 13}, (2003) 691--721.

\bibitem{DS}
A. Dhar and B.S. Shastry,
\emph{Bloch Walls and Macroscopic String States in Bethe's Solution
of the Heisenberg Ferromagnetic Linear Chain},
Phys. Rev. Lett., {\bf 85} (2000) 2813--2816.

\bibitem{DLS} F.~Dyson, E.H.~Lieb, and B.~Simon, \emph{Phase Transitions in
Quantum Spin Systems with Isotropic and Non-Isotropic Interactions},
J. Stat. Phys. \textbf{18} (1978) 335--383.

\bibitem{FSS} J.~Fr\"ohlich, B.~Simon, and T.~Spencer, \emph{Infrared bounds,
phase transitions and continuous symmetry breaking},
Comm. Math. Phys. \textbf{50} (1976) 79--95.

\bibitem{GW}
C.T. Gottstein and R.F. Werner, \emph{Groundstates of the $q$-deformed
Heisenberg ferromagnet},
preprint 1994, \mbox{cond-mat/9501123}.

\bibitem{GS}
L.-H. Gwa and H. Spohn,
\emph{Bethe solution for the dynamical-scaling exponent of the noisy
Burgers equation},
Phys. Rev, {\bf A46}, (1992) 844--854.

\bibitem{KLS} T.~Kennedy, E.H.~Lieb, and S.~Shastry, \emph{
Existence of N\'eel Order in Some Spin 1/2 Heisenberg Antiferromagnets},
J. Stat. Phys. \textbf{53} (1988) 1019--1030.

\bibitem{KLS2} T.~Kennedy, E.H.~Lieb, and S.~Shastry, \emph{
The XY Model has Long-Range Order for all Spins and all Dimensions Greater
than One},  Phys. Rev. Lett. \textbf{61} (1988) 2582--2584.

\bibitem{KN1}
T.~Koma and B.~Nachtergaele, \emph{The spectral gap of the ferromagnetic {XXZ}
chain}, Lett. Math. Phys. \textbf{40} (1997), 1--16, \mbox{cond-mat/9512120}.

\bibitem{KN3}
T.~Koma and B.~Nachtergaele, \emph{The complete set of ground states of the
ferromagnetic {XXZ} chains},
Adv. Theor. Math. Phys. \textbf{2} (1998), 533--558, \mbox{cond-mat/9709208}.

\bibitem{Lieb67a}
E.H.~Lieb, \emph{The residual entropy of square ice}, Phys. Rev. \textbf{162}
(1967), 162--172.

\bibitem{L} E.H.~Lieb, \emph{Two Theorems on the Hubbard Model},
Phys. Rev. Lett. \textbf{62} (1989) 1201--1204.

\bibitem{LM}
E.H.~Lieb and D.~Mattis, \emph{Ordering energy levels of interacting spin
systems},
J. Math. Phys. \textbf{3} (1962), 749--751.

\bibitem{LN} E.H.~Lieb and B.~Nachtergaele, \emph{The Stability of the Peierls
Instability for Ring Shaped Molecules},
 Phys. Rev. B \textbf{51} (1995) 4777--4791.

\bibitem{LS} E.H.~Lieb and P.~Schupp, \emph{Ground State Properties of a
Fully Frustrated Quantum Spin System},
Phys. Rev. Lett. \textbf{83}, (1999) 5362--5365.

\bibitem{MN} N.~Macris and B.~Nachtergaele, \emph{On the flux phase
conjecture at half-filling: an improved proof},
J. Stat. Phys. \textbf{85} (1996), 745--761.

\bibitem{Mor}
B. Morris,
\emph{The mixing time for simple exclusion},
preprint.

\bibitem{NS}
B. Nachtergaele and L. Slegers,
\emph{Construction of Equilibrium States for One-Dimensional Classical
Lattice Systems},
Il Nuovo Cimento, {\bf 100 B}, (1987) 757--779.

\bibitem{NSt}
B. Nachtergaele and S. Starr,
\emph{Droplet states in the XXZ Heisenberg Chain},
Commun. Math. Phys., {\bf 218}, (2001) 567-607.

\bibitem{NP}
E. Jord\~{a}o Neves and J. Fernando Perez, \emph{Long range order
in the ground state of two-dimensionsal antiferromagnets}, Phys.
Lett.   {\bf 114A}, (1986) 331--333.

\bibitem{Simon}
B. Simon, \emph{The Statistical Mechanics of Lattice Gases, Volume 1},
Princeton University Press, 1993.

\bibitem{Speer}
E.R. Speer,
\emph{Failure of reflection positivity in the quantum Heisenberg ferromagnet},
Lett. Math. Phys. {\bf 10}, (1985) 41--47.

\bibitem{TL} H.N.V.~Temperley and E.H.~Lieb, \emph{Relations between the
`percolation' and `colouring' problem and other graph-theoretical problems
associated with regular planar lattices: some exact results for the
`percolation' problem}, Proc. Roy. Soc., {\bf A322} (1971), 252--280.

\end{thebibliography}
\end{document}